\documentclass[reprint,amsmath,amssymb,aps,prapplied,floatfix]{revtex4-2}
\usepackage{graphicx}
\usepackage{dcolumn}
\usepackage{bm}
\usepackage{enumitem}
\usepackage[normalem]{ulem}
\usepackage[utf8]{inputenc}
\usepackage{xcolor}
\usepackage{upgreek}
\usepackage{textcomp}
\usepackage[hidelinks,colorlinks=true,linkcolor=blue,citecolor=blue, urlcolor=blue]{hyperref}
\begin{document}
\preprint{APS/123-QED}
\title{Event shapes and Inclusive Hadron Spectra at FCC-ee energies}
\author {Philip Mathew}
\email{joemat007@gmail.com}
\affiliation{CERN, 1211 Geneva 23, Switzerland}
\author {R. Aggarwal}
\email{ritu.aggarwal1@gmail.com}
\affiliation{USAR and USBAS, Guru Gobind Singh Indraprastha University, East Delhi Campus, 110092, India}
\author {M. Kaur}
\email{manjit@pu.ac.in}
\affiliation{Department of Physics, Panjab University, Chandigarh 160014, India\\
Department of Physics, Amity University, Punjab, Mohali 140306, India
}

\date{\today}

\begin{abstract}
We analyze hadronic final states of $e^+e^-$ annihilation through event shape observables, Thrust and C-parameter, and inclusive hadron spectra at the planned center-of-mass (c.m.)~energies of the Future Circular Electron-Positron Collider (FCC-ee). Collision data is produced using Monte Carlo event generation in PYTHIA at 91.2, 160, 240, and 365 GeV. Distortions of event shapes due to initial-state photon radiation and background decays of Z pairs, W pairs, top-quark pairs, and Higgs bosons are investigated. An extraction of the strong coupling $\alpha_{\text{s}}$ is performed by fitting event shape distributions to perturbative QCD predictions at next-to-next-to-leading-order (NNLO) accuracy, and the sources of systematic uncertainties at high c.m.~energies are discussed. Soft gluon dynamics is examined through charged particle multiplicities and momentum distributions, and energy evolution of mean values is compared with prior experimental results. The inferences from this phenomenological study provide a reference to QCD studies at future high-energy $e^+e^-$ colliders.
\end{abstract}

\maketitle
\section{Introduction}
The strong interaction has been successfully modeled by Quantum Chromodynamics (QCD), the non-abelian SU(3) gauge theory describing the forces binding fermions into composite hadrons~\cite{Ellis:1996}. The strength of this force between quarks mediated by gluons is parameterized by the strong coupling constant ($\alpha_\text{s}$)~\cite{Deur:2016}. Precise determinations of $\alpha_\text{s}$ are a fundamental ingredient in perturbative QCD predictions~\cite{Enterria:2024}. Uncertainties in the coupling carry over as theory errors in other sectors, including Higgs processes~\cite{Anastasiou:2016,Demartin:2010}, electroweak precision quantities~\cite{Heinemeyer:2021, Marc:2025}, and top-quark properties~\cite{Hoang:top}. The present relative uncertainty on $\alpha_\text{s}$ of 0.76\% as per PDG~\cite{PDG-review} has been made possible by decades of progress in accelerator-based experiments involving hadron-hadron~\cite{UA1:1981, Bala:2016, Kodolova:2025}, lepton-lepton~\cite{Stone:1995}, and lepton-hadron~\cite{Abramowicz:1999, shitsev_collider} collisions. The success of electron-positron colliders for QCD studies, in particular is largely due to a decoupled electroweak initial state allowing direct access to strong force effects~\cite{Kluth_2006}.

Although the wealth of data generated prior to and at the Large Electron-Positron collider~(LEP) has facilitated significant validity tests of QCD models, there still remains the need to constrain current models at higher precision, and develop new models to explain the non-perturbative regime~\cite{Ploerer:2022, fcc_cdr, Tor:2013}. Advancing the energy frontier from the LEP to the Large Hadron Collider~(LHC) shifted the experimental focus from precision measurements to discovery-oriented studies where higher momentum transfers complicate the isolation of softer QCD signatures~\cite{Gross:2023}. The proposed leptonic Future Circular Collider~(FCC-ee) is thus the next major opportunity to progress the strong sector. The latest design report specifies a 15-year plan over a wide range of center-of-mass (c.m.)~energies from just below the Z-pole to the W-boson pair threshold, the Higgsstrahlung maximum, and up to just above the top-quark pair threshold~\cite{fcc_fsr}. The QCD program aims to determine $\alpha_\text{s}$ at a 0.1\% accuracy~\cite{Enterria_alphas}, investigate perturbative parton radiation through jet rates and event shapes, and refine existing phenomenological models of non-perturbative parton hadronisation~\cite{Enterria_radiation}. 

Despite the exciting prospects of this venture, higher collision energies bring an extended set of challenges such as new sources of systematic uncertainties in $\alpha_\text{s}$ determinations. Enhanced initial state radiation (ISR) via QED reduces the effective energy of a large fraction of annihilations, requiring radiative cuts to ensure monochromatic results at the cost of statistics. Additional electroweak backgrounds distort the signal, requiring subtractive corrections further increasing errors. As a consequence, careful consideration of systematic uncertainties will be crucial in leveraging the latest advancements in theoretical models of event shape observables~\cite{Duca:2016,Abbate:2011,Hoang:2015}. The impacts of these challenges need to be studied in detail to facilitate fruitful experimental outcomes. To this end, we present a comprehensive study of hadronic $e^+e^-$ events in PYTHIA 8.313 through event shape observables and inclusive hadron spectra. This includes corrections for radiative and background distortions, a next-to-next-to-leading-order (NNLO) determination of $\alpha_{\text{s}}$ with Thrust and C-parameter distributions, and the inclusive dynamics of hadron production probed through charged multiplicity and momentum distributions. 

The paper begins with a brief review of QCD processes in electron-positron annihilation followed by the definitions of the event shape observables used in this study. Section III covers the Monte Carlo generation in PYTHIA and the validation of results with experimental data from the LEP. Section IV studies the influence of ISR and backgrounds on observables illustrating the importance of rethinking correction strategies at the FCC-ee. Section V presents an extraction of $\alpha_\text{s}$ by fitting event shapes to theory predictions, a description of the employed fitting procedure, and a discussion on sources of systematic errors at the FCC-ee. Finally, Section VI examines inclusive hadron spectra and compares the energy evolution of their mean values with previous data. 

\section{Event shape observables}
\begin{figure}[b]
\includegraphics[width=1\linewidth]{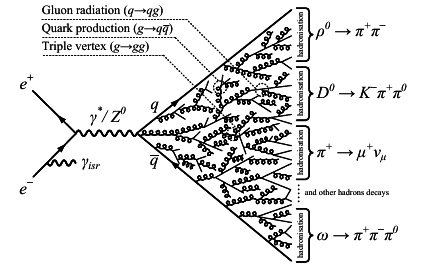}
\caption{\label{fig:01}Illustration of a hadronic $e^{+}e^{-}\rightarrow \gamma/Z\rightarrow q\overline{q}$ event. Hadrons and decay channels shown are only representative.}
\end{figure}

The typical structure of an annihilation event through the channel $e^{+}e^{-}\rightarrow \gamma^*/Z^0$ is illustrated in Fig.~\ref{fig:01}. At first, the incoming leptons may radiate a photon (ISR) before annihilating into a virtual photon or a Z-boson. The boson decays into a $q\overline{q}$ pair with sufficient energy to begin radiating gluons ($q\rightarrow qg$). The gluon can either become another quark pair ($g\rightarrow q\overline{q}$) or continue radiating gluons ($g\rightarrow gg$). Repeated continuation of these three QCD processes forms a cascading shower of quarks and gluons gradually approaching lower energy regimes. Due to the inverse running of $\alpha_{\text{s}}$ with energy, the partons experience a growing strong force pulling them together until they are rapidly forced by confinement into colorless mesons or baryons. These newly formed hadrons have definite lifetimes, short or long, and eventually decay into the stable hadrons that form the experimental observations inside the detector. The so called \textit{hard} phase is where parton production occurs with sufficiently high momentum transfers to be directly calculated with perturbation theory. The \textit{soft} phase is where hadronisation occurs and can be emulated through phenomenological models or power corrections to event shapes~\cite{Bambah_generators, Skands_2005}. 

QCD predicts that hadronic final states can have either a 2-jet structure of a quark and antiquark ($e^+e^-\rightarrow q\overline{q}$) or a multi-jet structure with an additional radiated gluon ($e^+e^-\rightarrow q\overline{q}g$). While the former contains large non-perturbative hadronisation corrections, the hard gluon in the latter introduces a perturbatively calculable deviation. The dynamics of this deviation can be probed by quantifying the geometrical distributions or `shapes' of momenta within a single event. This is achieved by constructing \textit{event shape observables} as linear sums of particle momenta that are intrinsically sensitive to hard gluon radiation. The extent to which these event observables deviate from the 2-jet scenario, when compared to corresponding theoretical predictions, allows a direct determination of the strong coupling. Importantly, this fitting procedure requires such observables to be infrared-safe and collinear-safe by ensuring insensitivity to soft gluon emissions and splittings at small angles, respectively. 

The event shapes that were used in QCD studies at the LEP commonly included the Thrust, C-parameter, Total and Wide jet broadening, and Heavy jet mass~\cite{Kluth_2006}. The thrust T measures the jet-like nature of an event by selecting an axis $\vec{n}$ that maximizes the projections of particle momenta $p_{i}$ and computing the expression:
\begin{equation}
\text{T} = \max_{\vec{n}} \left( \frac{\sum_i |\vec{p}_i \cdot \vec{n}|}{\sum_i |\vec{p}_i|} \right)
\end{equation}

Exact alignment of $p_{i}$ with $\vec{n}$ gives T=1 for 2-jet events and decreases with additional gluon radiation to a minimum of T=0.5 for a perfectly spherical event. The quantity (1-T) is used instead since it vanishes in the 2-jet limit and is better suited for theoretical expansions. The \text{C}-parameter measures the angular distribution of particle momenta by deriving eigenvalues $\lambda_i$ of the linear momentum tensor $\Theta^{ij}$ and computing the expression:
\begin{equation}
\Theta^{ij} = \frac{1}{\sum_k |\vec{p}_k|} \sum_k \frac{p_k^i p_k^j}{|\vec{p}_k|}, \qquad i,j = 1,2,3
\end{equation}
\begin{equation}
\text{C} = 3(\lambda_1 \lambda_2 + \lambda_2 \lambda_3 + \lambda_3 \lambda_1)
\end{equation}

Here, the term $p_k^i$ denotes the $i\textsuperscript{th}$ component of the momentum $\vec{p}_k$ of the $k\textsuperscript{th}$ particle of the event. Just as for (1-T), a 2-jet topology gives C = 0 and increases with radiation to a maximum of C = 1. Although both observables are quantitatively constructed from similar geometries of hadronic final states, they qualitatively highlight different aspects of QCD physics. Thrust is more sensitive to perturbative hard gluon emissions while the C-parameter also responds significantly to non-perturbative wide-angle soft radiation that adds event isotropy. This complementarity is the reason for selecting these observables to make this study focused yet comprehensive. The presented work focuses on studying these two variables in detail at the proposed c.m. energies of the FCC-ee. 

\clearpage
\section{Event Generation}
\begin{figure}[b]
\includegraphics[width=1\linewidth]{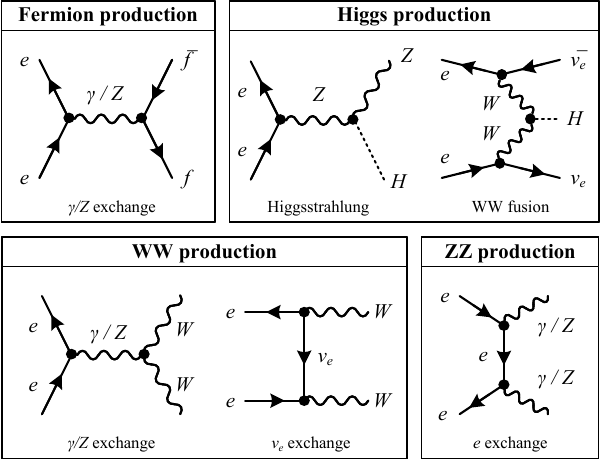}
\caption{\label{fig:02}Dominant annihilation processes at the FCC-ee.}
\end{figure}

\begin{figure}[b]
\includegraphics[width=1\linewidth]{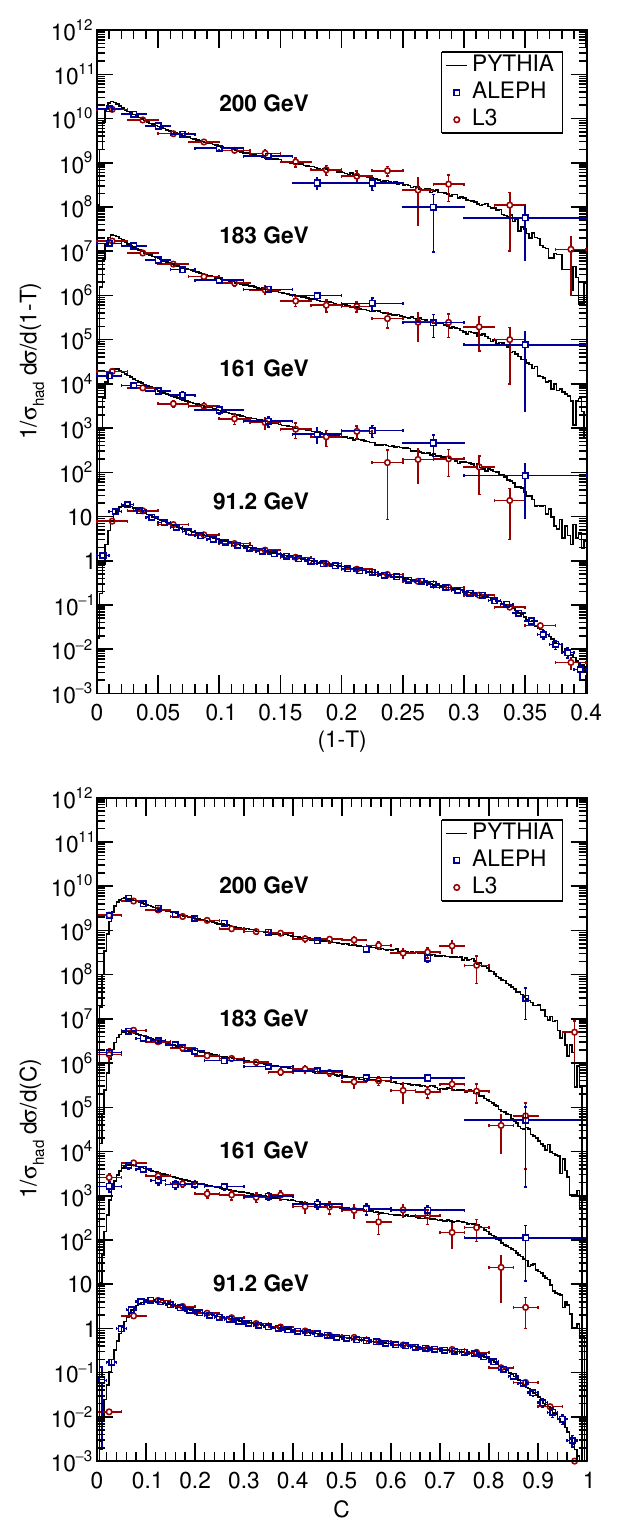}
\caption{\label{fig:03}Comparison of normalized event shape distributions generated in PYTHIA 8.313 with experimental data from L3 (red circles) and ALEPH (blue squares). Each subsequent distribution is displaced by a factor of $10^3$ for presentation.}
\end{figure}

Samples of $5 \times 10^6$ events are generated in PYTHIA 8.313 at four c.m. energies: the Z-pole at 91.2 GeV, the WW threshold at 160 GeV, the ZH maximum at 240 GeV and the $\text{t}\overline{\text{t}}$ at 365 GeV. The simulations include only hadronic decays of the relevant electroweak processes illustrated in Fig.~\ref{fig:02}. The integrated luminosities computed from PYTHIA cross sections are summarized in Table~\ref{tab:1}. The estimated equivalent FCC-ee run time is calculated with the latest proposed physics run plan~\cite{fcc_fsr}. The default model for ISR in PYTHIA is turned on only for the results presented in Section IV.A~\cite{tor:2015}. Additionally, samples of $10^6$ events are generated at 161, 183, and 200 GeV, to compare simulated differential event shapes distributions against data collected at ALEPH~\cite{ALEPH:209gev} and L3~\cite{L3:209gev} in Fig.~\ref{fig:03}. Good agreement is observed in the 3-jet region ($0.1<(1-\text{T})<0.3$ and $0.3<\text{C}<0.7$). Deviations in the higher-jet regions (right of the 3-jet region) are due to limited statistics of experimental data. As a verification of hadronisation models, the obtained mean charged hadron multiplicities were within 0.5\% of experiment.

\begin{table}
\caption{\label{tab:1} Summary of process cross sections, the integrated luminosity, and the duration it represents in the FCC-ee run.}
\begin{ruledtabular}
\begin{tabular}{l|c|c|c|c}
Cross section $\sigma$ & 91.2 GeV & 160 GeV & 240 GeV & 365 GeV
\\ \hline
$ee(\gamma)\rightarrow \gamma/Z$           & 30.4 nb   & 151.1 pb  & 52.1 pb   & 21.1 pb
\\ 
$ee\rightarrow \gamma/Z$                   & 41.4 nb   & 38.6 pb   & 13.4 pb   & 5.4 pb
\\
$ee\rightarrow W^+W^-$                     & -         & 1.4 pb    & 7.7 pb    & 4.9 pb
\\
$ee\rightarrow ZZ$                         & -         & 0.2 pb    & 0.7 pb    & 0.4 pb
\\
$ee\rightarrow H\nu_e\overline{\nu_e}$     & -         & 0.1 fb    & 4.4 fb    & 21.9 fb
\\
$ee\rightarrow ZH$                         & -         & -         & 0.1 pb    & 52.7 fb
\\
$ee\rightarrow t\overline{t}$              & -         & -         & -         & 0.3 pb
\\ \hline
Sample size $N$                            & 5 M       & 5 M       & 5 M       & 5 M
\\
Luminosity $\mathcal{L}$                   & 0.12 fb$^{-1}$ & 124 fb$^{-1}$ & 230 fb$^{-1}$ & 455 fb$^{-1}$
\\
FCC-ee time                                & 1.5 min & 7 days & 36 days & 1 year
\end{tabular}
\end{ruledtabular}
\end{table}

\clearpage
\section{Challenges at high $\sqrt{\text{s}}$}
In this section, we analyze the effects of ISR and backgrounds on event shapes at c.m. energies of the FCC-ee.

\subsection{QED Radiation}
\begin{figure}
\includegraphics[width=1\linewidth]{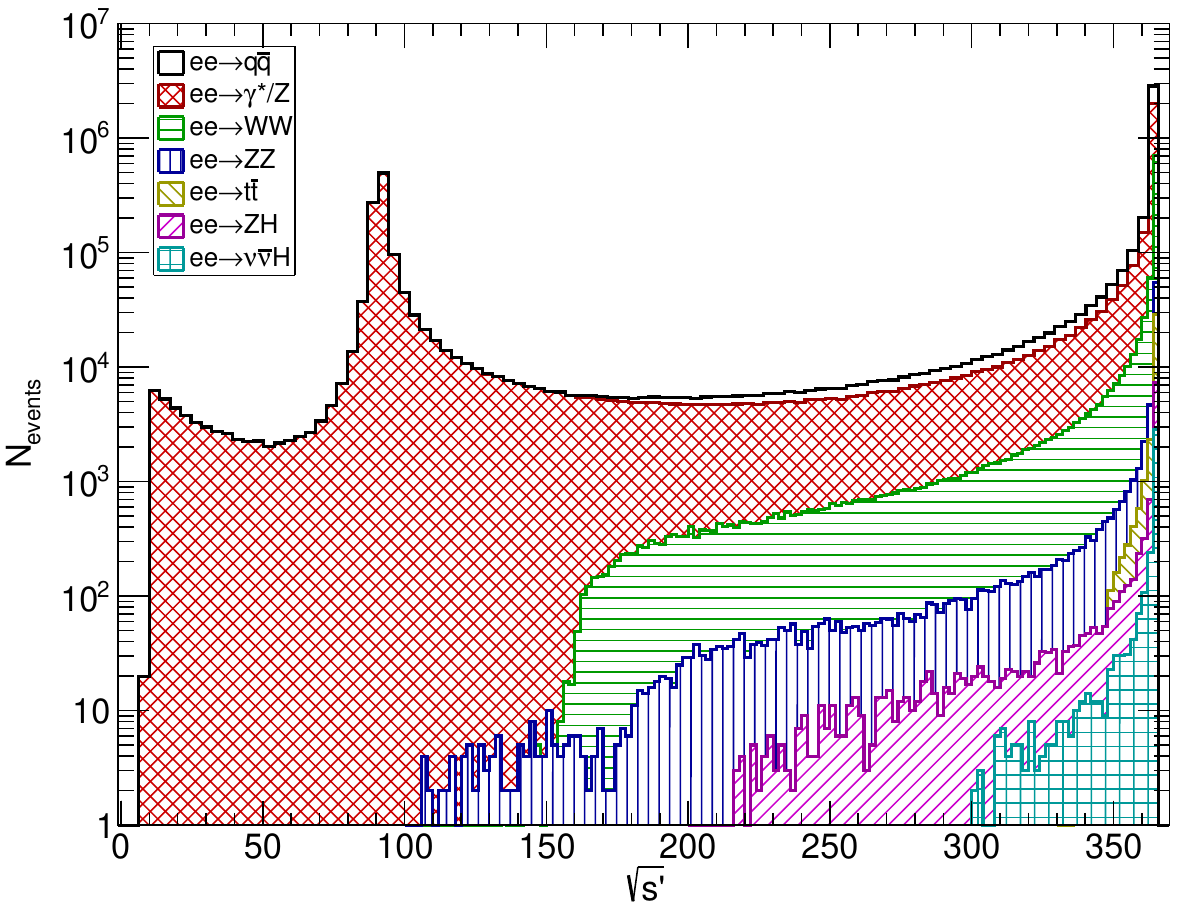}
\caption{\label{fig:04}Process-wise contributions to $\sqrt{s'}$ at $\sqrt{s}=365\;$GeV.}
\end{figure}

\begin{figure}[b]
\includegraphics[width=1\linewidth]{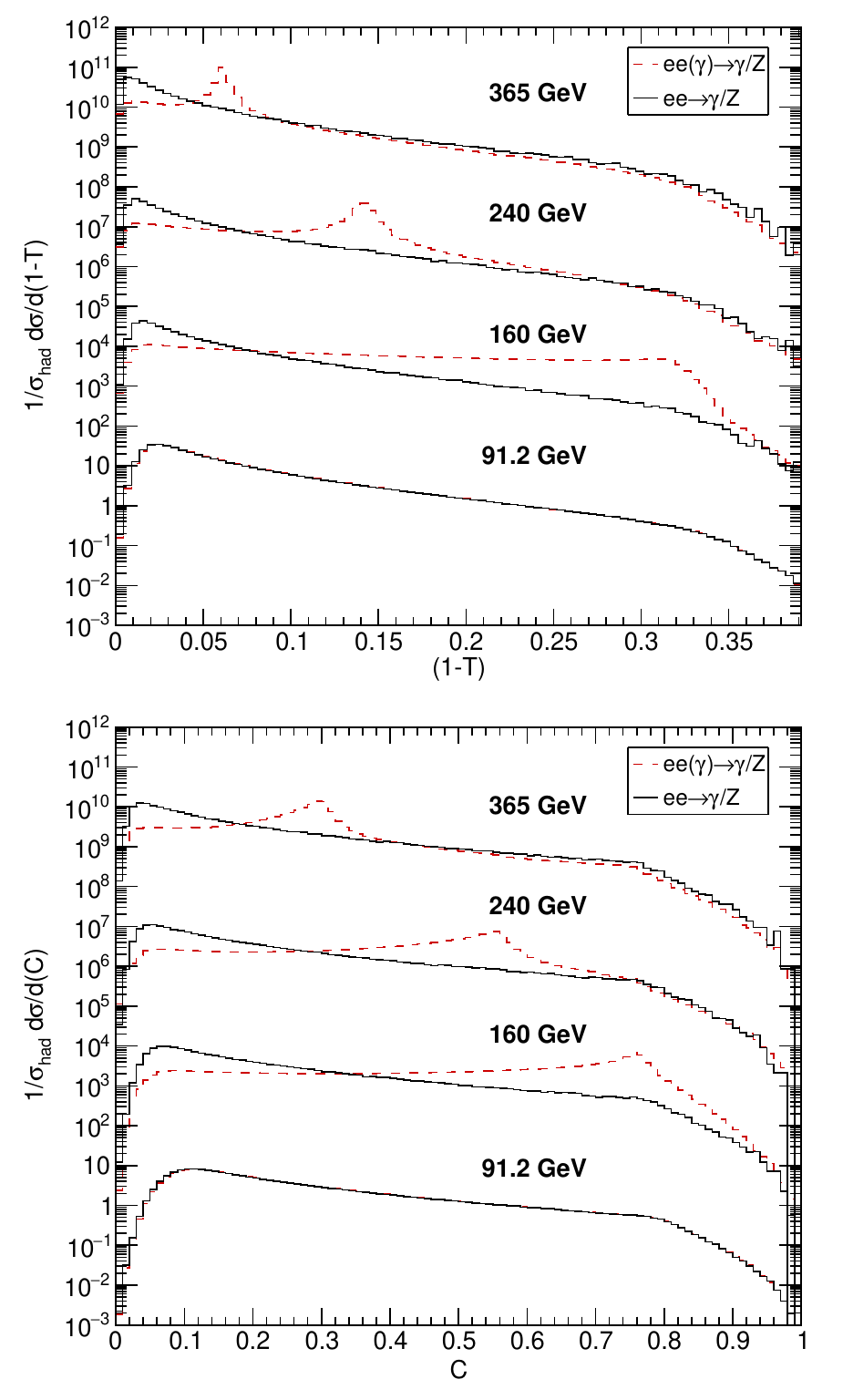}
\caption{\label{fig:05}Impact of ISR on event shapes (1-T) and C. Each subsequent distribution is displaced by a factor of $10^3$.}
\end{figure}

The radiation of photons from initial state leptons is a prominent feature of $e^+e^-$ colliders that lowers the effective c.m. energy of the remaining hadronic system $\sqrt{s'}$:
\begin{equation}
\sqrt{s'} \ =\ \sqrt{\ s \left( 1 - \frac{2E_\gamma}{\sqrt{s}} \right)}
\end{equation}
Where $E_\gamma$ is the energy of the ISR photon. At the LEP1, this enabled important energy-dependence studies of $\alpha_{s}$ below the Z-pole~\cite{schieck:2003}. However, the higher energies at the LEP2 experienced a large number of events with an $E_\gamma$ that returned the $\sqrt{s'}$ back to the $m_Z$ scale~\cite{L3:172gev}. These so called radiative returns is explained by integrating the differential full cross section of the $f_i \bar{f}_j \rightarrow Z \rightarrow f_{i}' \bar{f}_{j}'$ process into the Breit-Wigner form written as follows~\cite{PDG-review}:
\begin{equation}
\sigma(s') = \frac{12\pi}{m_Z^2} \frac{\Gamma_e \Gamma_f}{(s' - m_Z^2)^2 + m_Z^2 \Gamma_Z^2},
\end{equation}
Where $\Gamma_e,\Gamma_Z$ are the widths of the electron and Z-boson, respectively. The resonance at $\sqrt{s'}=m_Z^2$ leads to a boost of the $ee\rightarrow\gamma/Z$ cross section producing a characteristic peak at $\sqrt{s'}= 91.2$ GeV in the $\sqrt{s}= 365$ GeV plot shown in Fig.~\ref{fig:04}. Since the energy lost through ISR is predicted to be logarithmic with c.m. energy, a large number of radiative events can be expected at the FCC-ee~\cite{Altarelli:300671}. The impact of this radiation on event shapes is visualized in Fig.~\ref{fig:05}. The peaks observed at $\sqrt{s}=160,240,365$ GeV in (1-T)$\approx0.325,0.14,0.06$ and in $\text{C}\approx0.74,0.54,0.30$, respectively, can be attributed to the radiative returns~\cite{denner:2010}. These distortions will need to be filtered out through cuts on $\sqrt{s'}$ before performing $\alpha_{\text{s}}$ determinations. In Fig.~\ref{fig:05}, a tight cut ($\sqrt{s'}=1.00\sqrt{s}$) on the simulated distributions (red dashed) produces the clean signals (black solid). While this effectively removes the ISR distortions, only 4.7\% of the original statistics survived the cut. The survival fraction varies greatly with tighter cuts and c.m. energy as summarized in Table~\ref{tab:2}, pointing to a need to rethink ISR correction strategies at each FCC-ee level.

\begin{table}[t]
\caption{\label{tab:2} Summary of the survival fraction of $\gamma/Z$ events after employing various radiative cuts at each FCC-ee energy.}
\begin{ruledtabular}
\begin{tabular}{l|c|c|c|c}
Radiative cut & 91.2 GeV & 160 GeV & 240 GeV & 365 GeV
\\ \hline
$\sqrt{s'}<1.00\sqrt{s}$ & 43.2\% & 7.1\% & 5.7\% & 4.7\%
\\ 
$\sqrt{s'}\geq0.95\sqrt{s}$ & 99.8\% & 56.0\% & 49.9\% & 46.3\%
\\
$\sqrt{s'}\geq0.90\sqrt{s}$ & 99.9\% & 59.1\% & 52.4\% & 48.5\%
\\
$\sqrt{s'}\geq0.85\sqrt{s}$ & 100\% & 61.0\% & 53.9\% & 49.8\%
\end{tabular}
\end{ruledtabular}
\end{table}

\clearpage
\subsection{Electroweak backgrounds}
\begin{figure}[b]
\includegraphics[width=1\linewidth]{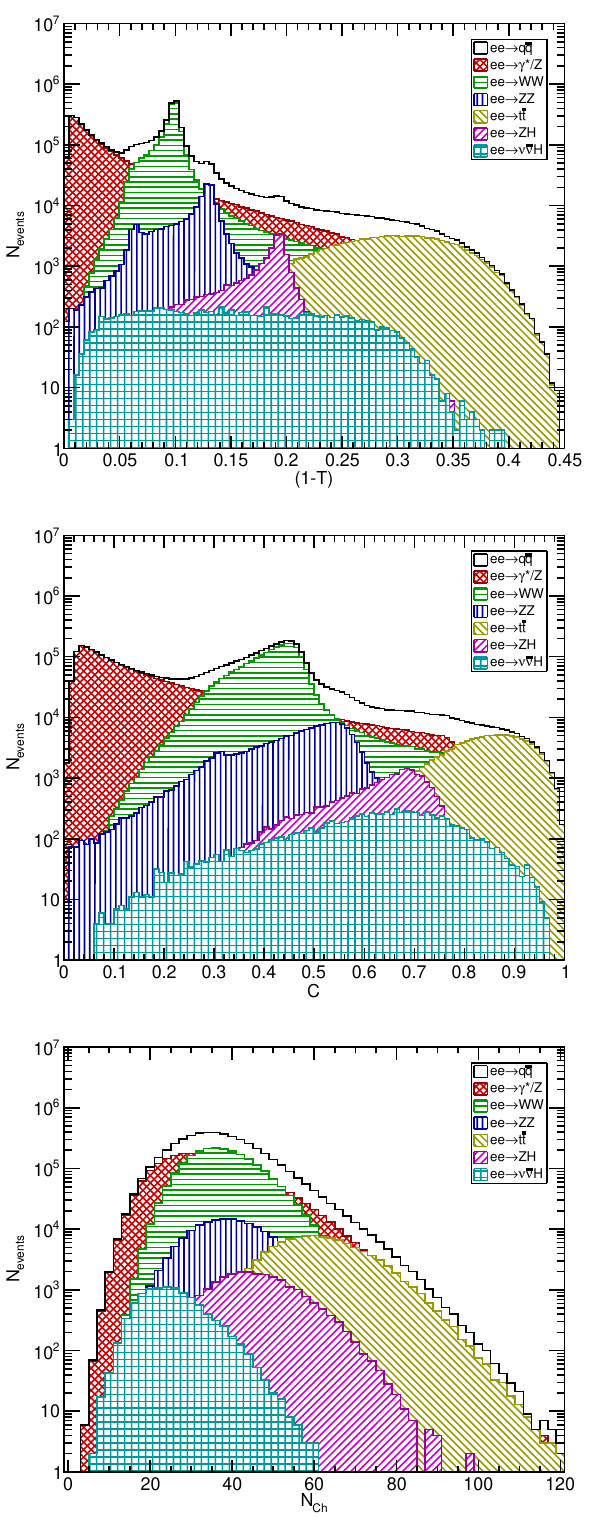}
\caption{\label{fig:06} Contribution of electroweak backgrounds to event shapes (1-T), C and charged multiplicity $\text{N}_{\text{ch}}$ at 365 GeV.}
\end{figure}

\begin{figure}[b]
\includegraphics[width=1\linewidth]{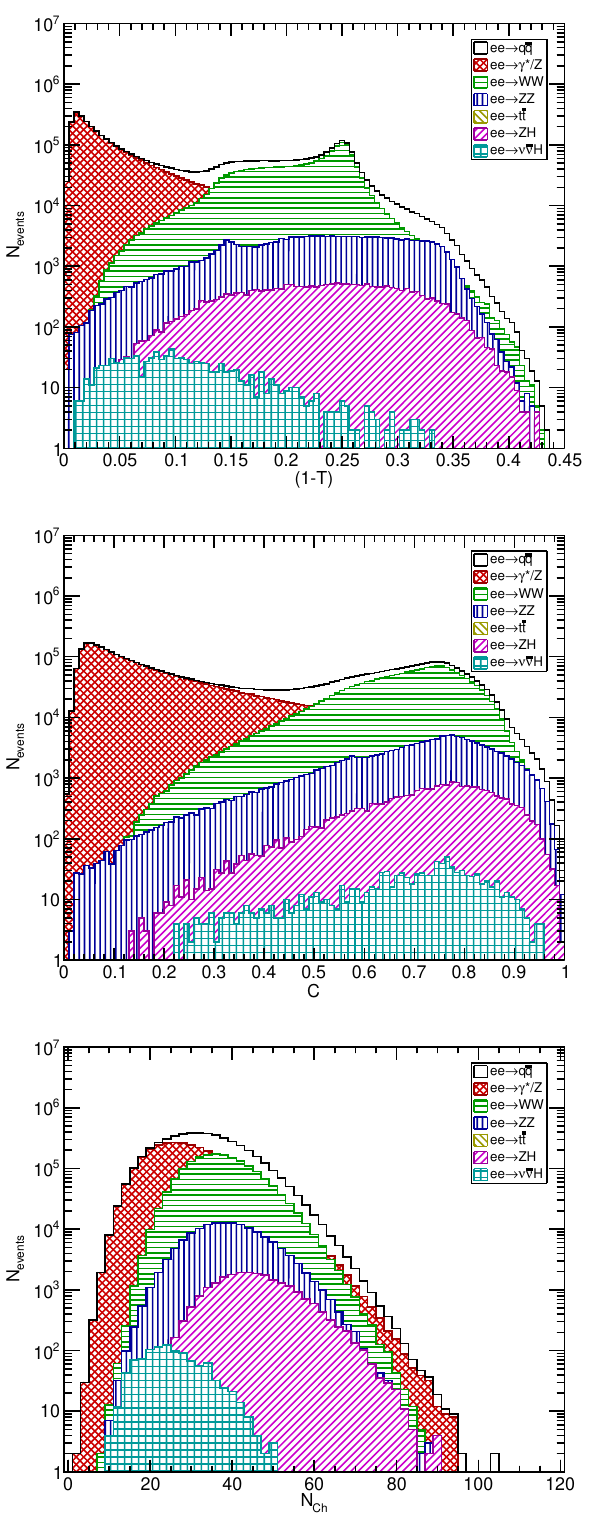}
\caption{\label{fig:07} Contribution of electroweak backgrounds to event shapes (1-T), C and charged multiplicity $\text{N}_{\text{ch}}$ at 240 GeV.}
\end{figure}

\begin{figure}[b]
\includegraphics[width=1\linewidth]{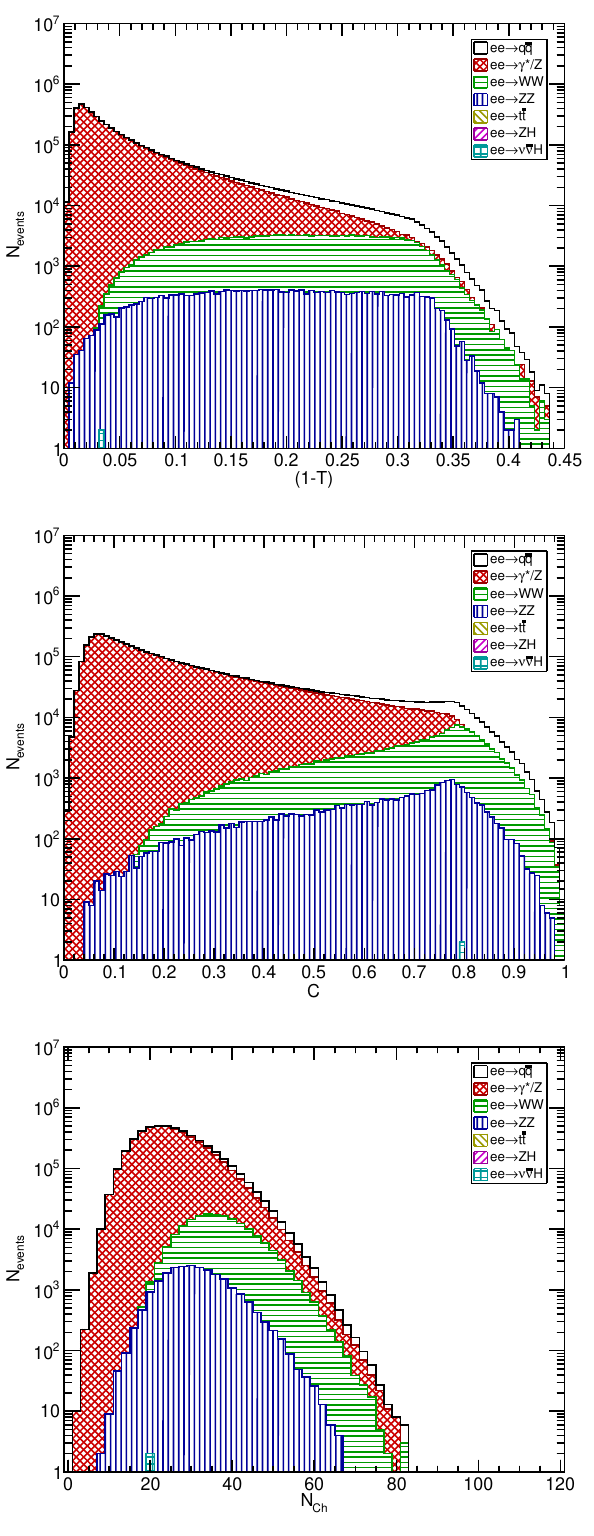}
\caption{\label{fig:08} Contribution of electroweak backgrounds to event shapes (1-T), C and charged multiplicity $\text{N}_{\text{ch}}$ at 160 GeV.}
\end{figure}

\begin{figure}[b]
\includegraphics[width=1\linewidth]{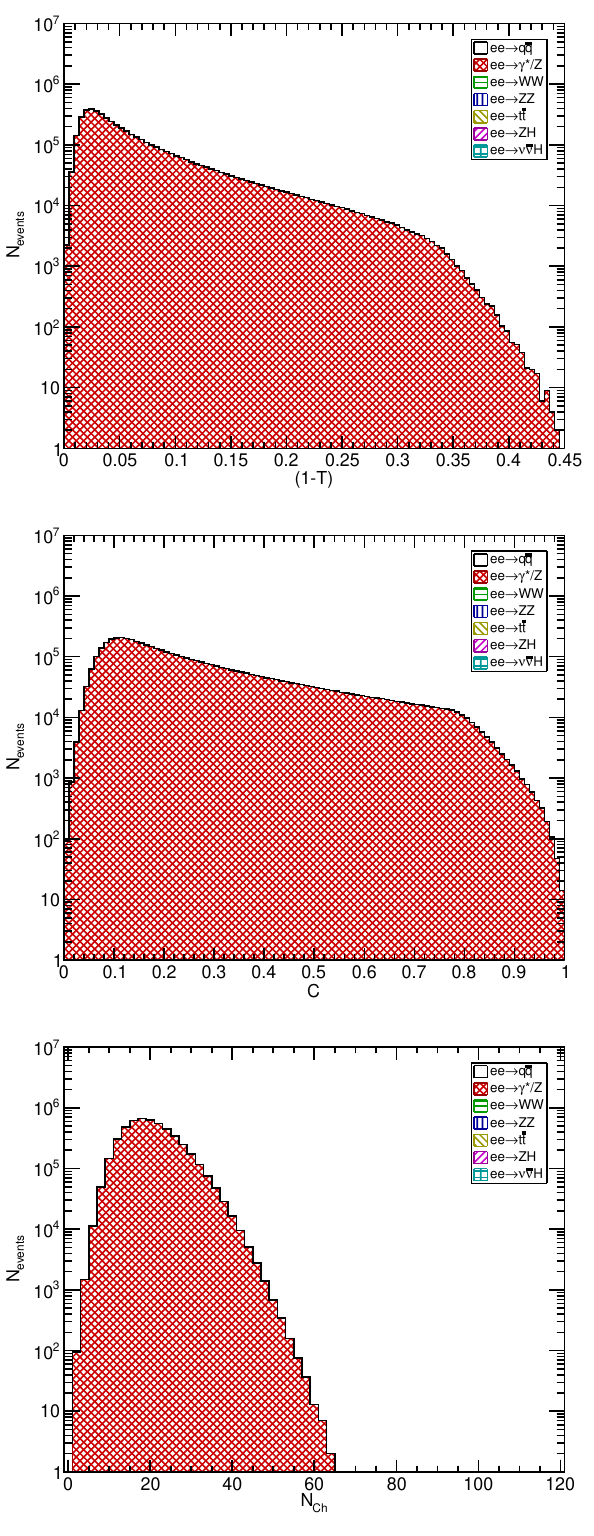}
\caption{\label{fig:09} Contribution of electroweak backgrounds to event shapes (1-T), C and charged multiplicity $\text{N}_{\text{ch}}$ at 91.2 GeV.}
\end{figure}

\begin{figure}[b]
\includegraphics[width=1\linewidth]{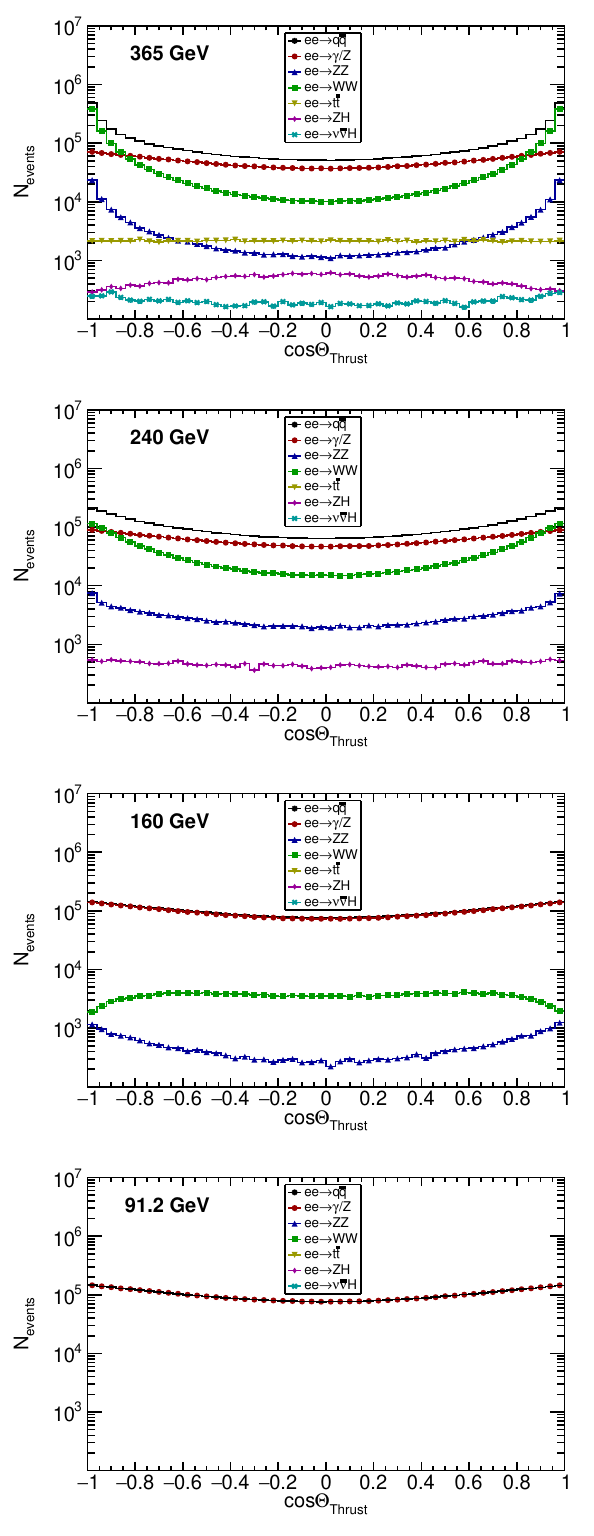}
\caption{\label{fig:10} Comparison of event thrust axis cos$\Theta_{\text{Thrust}}$ distributions of each electroweak background at FCC-ee energies.}
\end{figure}

Several electroweak processes are expected at FCC-ee energies constituting significant backgrounds to studies on the $ee \rightarrow \gamma/Z$ channel. The characteristic influence of each background is examined through event shape distributions, reflecting the geometry of final states, and through charged hadron multiplicity distributions, reflecting hadronization characteristics. On a sample of $5\times10^6$ events, the process-separated plots of these distributions are provided in Fig.~\ref{fig:06},~\ref{fig:07},~\ref{fig:08},~\ref{fig:09}. The production of W-pairs occurs above the $2m_W$ scale through s-channel $\gamma/Z$ and t-channel $\nu_e$ exchanges, while that of ZZ pairs occurs above the $2m_Z$ scale solely through the s-channel $\gamma/Z$ exchange. Both gauge boson predominantly decay into a quark-antiquark pair resulting in intricate 4-jet topologies ($q\overline{q}q\overline{q}$) in the final state. The resulting increase in event isotropy due to additional jets pushes (1-T) and C towards higher values as compared to the $\gamma/Z$ signal. Since the WW peak sits left of ZZ for both event shapes, this indicates that jets in WW events tend to be more collimated than those in ZZ events. Higgsstrahlung is an s-channel resonance producing back-to-back H and Z bosons. The corresponding event shapes are shifted further right to higher jet regions. On the other hand, WW fusion is a t-channel exchange without a resonance seen as a flat diffused event shape profile. The production of $\text{t}\overline{\text{t}}$ begins near the $2m_t$ scale via the $\gamma/Z$ exchange. Each heavy quark decays almost exclusively via the $t\rightarrow W^+b$ or the $\overline{t}\rightarrow W^-\overline{b}$ channels. The final state thus contains six quarks ($q\overline{q}q\overline{q}b\overline{b}$) breeding complex multi-jet topologies with significant gluon radiation. The enhanced spherical nature of such events contribute to high (1-T), C regions making it a distinctive background greatly sensitive to non-perturbative effects. The removal of electroweak backgrounds involves subtractive corrections with Monte Carlo generators. The expected reduction in $\gamma/Z$ statistics after such corrections is summarized in Table~\ref{tab:3}. Detector effects are also important to consider when studying distortions. Plots of thrust axis $\text{cos}\Theta_{\text{Thrust}}$ in Fig.~\ref{fig:10} illustrate how an imperfect angular acceptance would affect each electroweak process disproportionately. The wider coverage in the forward region expected at the FCC-ee as compared to detectors at the LEP could minimize these effects~\cite{IDEA:2025,Dam:2025,fcc_cdr}.

\begin{table}[t]
\caption{\label{tab:3} Summary of electroweak background statistics and survival fractions of $\gamma/Z$ events at each FCC-ee energy.}
\begin{ruledtabular}
\begin{tabular}{l|c|c|c|c}
Number of events & 91.2 GeV & 160 GeV & 240 GeV & 365 GeV \\ 
\hline
$ee\rightarrow \gamma/Z$                    & 5000000   & 4799181   & 3068124   & 2435489 \\
$ee\rightarrow W^+W^-$                      & -         & 176856    & 1758070   & 2248478 \\
$ee\rightarrow ZZ$                          & -         & 23954     & 149779    & 174812  \\
$ee\rightarrow H\nu_e\overline{\nu_e}$      & -         & -         & 1034      & 10052   \\
$ee\rightarrow ZH$                          & -         & -         & 22993     & 23717   \\
$ee\rightarrow t\overline{t}$               & -         & -         & -         & 107452  \\
\hline
Survival fraction                           & 100\%     & 96\%      & 61\%      & 49\%
\end{tabular}
\end{ruledtabular}
\end{table}
\clearpage
\section{Determination of $\alpha_{\text{s}}$}
This section briefly revisits the QCD formulation of event shape observables and performs an extraction of the strong coupling.~The fitting procedure is explained in detail and the results compared with prior LEP values.

\subsection{Theoretical description}
The fixed order differential distribution of an experimentally measured event shape observable $O$ can be expressed perturbatively up to NNLO accuracy as follows:
\begin{multline} \label{eq:bar}
\frac{O}{\sigma_{\text{had}}(\mu)} \cdot \frac{d\sigma(\mu)}{dO}
= \Bigg( \frac{\alpha_{\text{s}}(\mu)}{2\pi} \Bigg) \cdot \frac{ O \ d\overline{A}(\mu)}{dO}\\
+ \Bigg( \frac{\alpha_{\text{s}}(\mu)}{2\pi} \Bigg)^2 \cdot \frac{ O \ d\overline{B}(\mu)}{dO}
+ \Bigg( \frac{\alpha_{\text{s}}(\mu)}{2\pi} \Bigg)^3 \cdot \frac{ O \ d\overline{C}(\mu)}{dO}
\end{multline}

\noindent Where $\sigma_{\text{had}}$ is the total hadronic cross section and each term of coefficients $\overline{A},\overline{B},\overline{C}$ corresponds to the corrections at LO, NLO, and NNLO, respectively. The numerical parton-level computation of (\ref{eq:bar}) instead utilizes the form:
\begin{multline} \label{eq:unbar}
\frac{O}{\sigma_{\text{0}}(\mu)} \cdot \frac{d\sigma(\mu)}{dO}
= \Bigg( \frac{\alpha_{\text{s}}(\mu)}{2\pi} \Bigg) \cdot \frac{ O \ dA(\mu)}{dO}\\
+ \Bigg( \frac{\alpha_{\text{s}}(\mu)}{2\pi} \Bigg)^2 \cdot \frac{ O \ dB(\mu)}{dO}
+ \Bigg( \frac{\alpha_{\text{s}}(\mu)}{2\pi} \Bigg)^3 \cdot \frac{ O \ dC(\mu)}{dO}
\end{multline}

\noindent Where $\sigma_{\text{0}}$ is the LO cross section of the $ee\rightarrow$ hadrons process. The translation from (\ref{eq:bar}) to (\ref{eq:unbar}) is then simply:
\begin{align}
\bar{A}_0 &= A_0 \label{eq:ao}\\
\bar{B}_0 &= B_0 - A_{\text{tot}} A_0 \label{eq:bo}\\
\bar{C}_0 &= C_0 - A_{\text{tot}} B_0 - \left( B_{\text{tot}} - A_{\text{tot}}^{2} \right) A_O
\end{align}

\noindent And coefficients $A_{\text{tot}}$ and $B_{\text{tot}}$ are defined as follows:
\begin{align*}
A_{\text{tot}} &= \frac{3\,(N_c^2 - 1)}{4N_c} \\
B_{\text{tot}} &= \frac{N_c^2 - 1}{8N_c} \left[ \left( \tfrac{243}{4} - 44\,\zeta_3 \right) N_c
+ \tfrac{3}{4N_c} + \left( 8\,\zeta_3 - 11 \right) N_f \right]
\end{align*}

\noindent Here, we select the $N_c=3$ as the number of quark colors and the $N_f=~5$ as the number of active quark flavors. Considering massless quarks, the ratio of cross sections,

\begin{equation}
\frac{\sigma_{\text{had}}}{\sigma_{0}}
= 1 + \frac{\alpha_{\text{s}}}{2\pi} A_{\text{tot}}
+ \left(\frac{\alpha_{\text{s}}}{2\pi}\right)^{2} B_{\text{tot}}
+ \dots \ \mathcal{O}(\alpha_{\text{s}}^{n})
\end{equation}

\noindent contains canceling electroweak coupling factors.~A correction for the bottom quark is required when the $b\overline{b}$ events are considerable.~These effects scale with $m_b^2/Q^2$, accounting to nearly 1\% at 91.2 GeV and 0.2-0.3\% at 200 GeV~\cite{Barate_2000}.~The correction is derived by subtracting coefficients in (\ref{eq:ao}) and (\ref{eq:bo}) for massless and massive scenarios as calculated in~\cite{Chetyrkin_1979}.~The top quark mass also needs corrections at energies above the $\text{t}\overline{\text{t}}$ threshold. Both these contributions have not been included in the current study.

\subsection{Fitting procedure}
\begin{figure}
\includegraphics[width=1\linewidth]{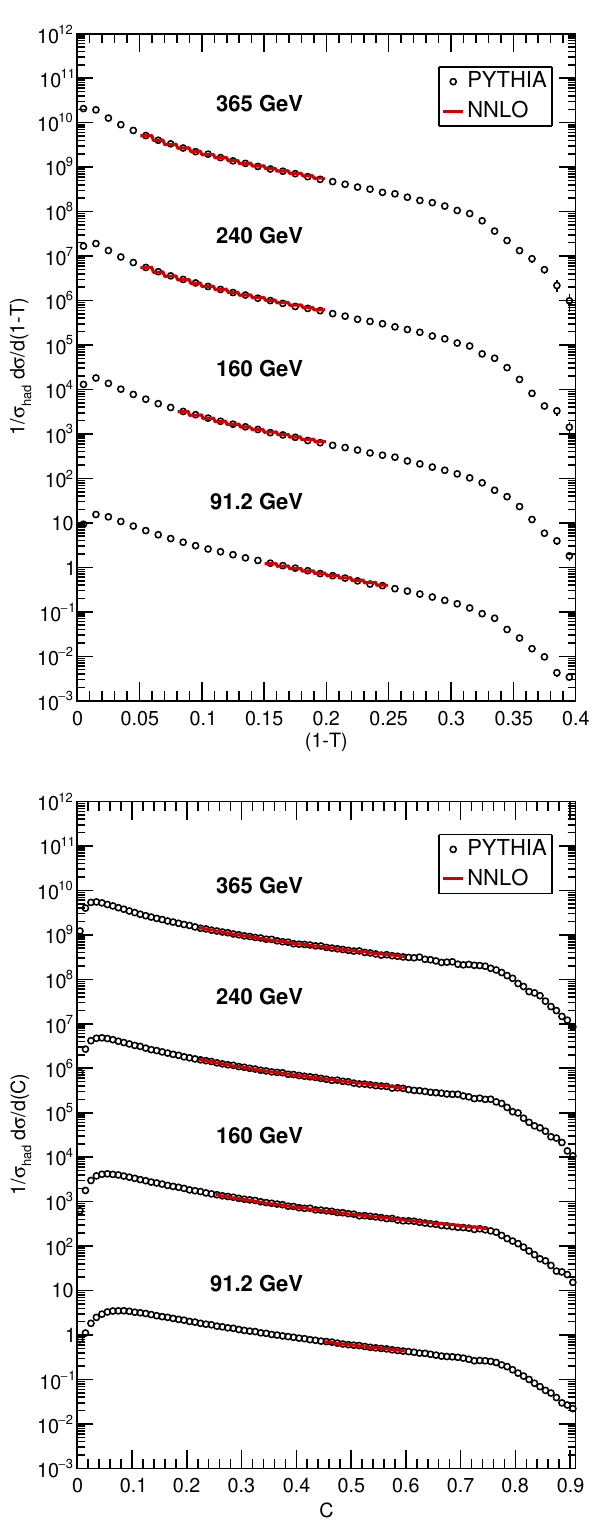}
\caption{\label{fig:12}NNLO fits to event shapes at FCC-ee energies denoted by red lines and black circles, respectively. Each distribution is scaled by a factor of $10^3$ solely for presentation.}
\end{figure}

Samples of $5 \times 10^6$ events are generated in PYTHIA without hadronisation and without ISR at each FCC-ee energy. The $ee \rightarrow \gamma/Z$ channel is isolated for this study. Event shape observables are then computed directly from the four-momenta of all final-state particles without explicit jet clustering. NNLO predictions are constructed using coefficients from Weinzierl~\cite{Weinzierl_2009}. Simulated event shape distributions are fitted using the MINOS routine of the MINUIT minimization framework in ROOT~\cite{James:2004}. The extracted $\alpha_{\text{s}}$ values from each observable are averaged with equal weights and the results summarized in Table~\ref{tab:4}. The selected fit ranges shown in Fig.~\ref{fig:12} are inspired from~\cite{ALEPH:209gev} and correspond to the 3-jet region where perturbative predictions are most reliable and hadronisation corrections remain minimal. Extending the window, especially to the left of this region, would require resummed logarithmic corrections to the NNLO predictions. The lower limits of most fits were shifted to slightly higher values for reasonable results, potentially due to the limitations of fixed-order theory predictions in non-perturbative modeling. In all fits, the $\chi^2/dof$ is in between 1 and 1.8, indicating good agreement between simulation and theory across energies. The value at 91.2 GeV from the (1-T) and C fits are consistent within approximately 1\% of the NNLO result in~\cite{Dissertori_2009} thereby validating reliability of the fitting procedure. The deviation of $\alpha_{\text{s}}$ between event shapes is $\leq$0.6\% at all c.m. energies.

\subsection{Uncertainties at FCC-ee}
\begin{figure}
\includegraphics[width=1\linewidth]{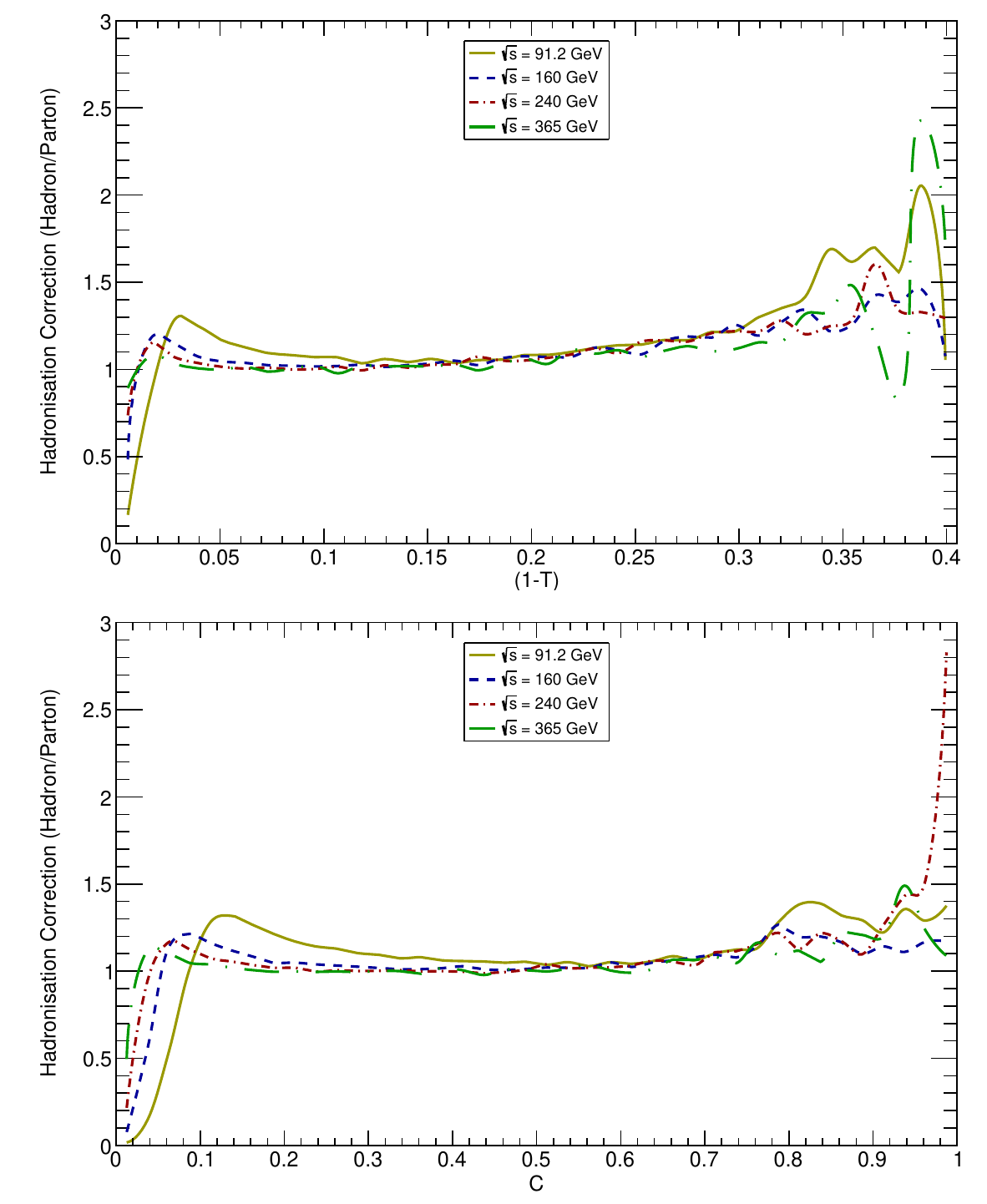}
\caption{\label{fig:11}Hadronisation correction factors to the (1-T) and C event shapes at FCC-ee energies computed in PYTHIA 8.313.}
\end{figure}

The large luminosities planned for the FCC-ee will limit the precision of $\alpha_{\text{s}}$ measurements to primarily systematic rather than statistical uncertainties~\cite{fcc_cdr}. At LEP, these systematic errors decreased with $\sqrt{s}$ from approximately 8\% at 91.2 GeV to 5\% at 206 GeV~\cite{L3:209gev}, and a similar trend is expected at higher energies. Theoretical models of event shapes have improved through the matching of resummed logarithmic terms to fixed-order predictions, improving reliability near the 2-jet region further reducing uncertainties~\cite{Jones_2003}. The latest N\textsuperscript{3}LL+N\textsuperscript{3}LO precision has been achieved for (1-T) by Abbate \textit{et al.}~\cite{Abbate:2011} and for C by Hoang \textit{et al.}~\cite{Hoang:2015}. The use of hadronisation correction factors on experimental measurements to relate parton-level and hadron-level data introduce an additional source of systematic uncertainty. These have been computed in PYTHIA for each FCC-ee energy and show a reduced impact at higher c.m. energies in Fig.~\ref{fig:11}~\cite{Biebel1999}. The deviation in hadronisation corrections in the fit range between the PYTHIA and HERWIG event generators has been 
studied at LEP energies to be in the range of 0.5--3\% \cite{Dissertori_2009}. The resulting generator-dependent spread in $\alpha_{\text{s}}$ values from fitting to ALEPH data was 0.001 ($\approx$ 0.8\%) for Thrust and C-parameter \cite{Dissertori_2009}. The removal of distortions from ISR and electroweak backgrounds discussed in Section IV will add to the overall systematic error budget. Extending this study in DELPHES would help model detector-related errors~\cite{DELPHES}. Achieving tight control over systematic uncertainties will hence be crucial for $\alpha_{\text{s}}$ studies at FCC-ee.

\begin{table}
\caption{\label{tab:4} Strong coupling $\alpha_{\text{s}}$ obtained from fitting NNLO theory to PYTHIA event shapes at FCC-ee energies.}.
\begin{ruledtabular}
\begin{tabular}{c|c|c|c|c|c}
Shape & Result & 91.2 GeV & 160 GeV & 240 GeV & 365 GeV \\
\hline
& $\alpha_{S}$ & 0.1284 & 0.1190 & 0.1127 & 0.1064 \\
(1-T) & $\chi^2/dof$ & 9.3/9 & 18.8/11 & 24.7/14 & 14.4/14 \\
& Range & 0.15--0.25 & 0.08--0.20 & 0.05--0.20 & 0.05--0.20 \\
& Error & $\pm$0.0001 & $\pm$0.0002 & $\pm$0.0002 & $\pm$0.0002 \\
\hline
& $\alpha_{S}$ & 0.1284 & 0.1183 & 0.1120 & 0.1058 \\
C & $\chi^2/dof$ & 25.9/14 & 84.8/49 & 65.6/37 & 58.8/37 \\
& Range & 0.45--0.60 & 0.25--0.75 & 0.22--0.60 & 0.22--0.60 \\
& Error & $\pm$0.0001 & $\pm$0.0001 & $\pm$0.0002 & $\pm$0.0002 \\
\hline
Average & $\alpha_{S}$ & 0.1284 & 0.1186 & 0.1123 & 0.1061 \\
& Error & $\pm$0.0001 & $\pm$0.0001 & $\pm$0.0002 & $\pm$0.0002 \\
\end{tabular}
\end{ruledtabular}
\end{table}

\section{Inclusive hadron spectra}
Inclusive observables refer to energy-dependent measurements that sum over all final state hadrons irrespective of process or topology. Capturing global properties of hadronic final states provide a probe to the average behavior of parton fragmentation and hadronisation. This assumed correlation that hadron-level dynamics reflects those at the parton-level scaled by a constant is termed Local Parton--Hadron Duality (LPHD)~\cite{azimov_duality, azimov_hump}. We examine here the energy evolution of two inclusive spectra of charged hadrons, the multiplicity and the momenta.

\subsection{Charged hadron multiplicity}
The total hadron multiplicity $\text{N}_{\text{tot}}$ represents the cumulative outcome of all non-perturbative dynamics.~Since only charged hadrons can be detected experimentally, the corresponding charged hadron multiplicity $\langle \text{N}_{\text{ch}}\rangle$ is the typically measured observable. The correction factor $\text{K}_{\text{ch}}$ for neutral contributions is found to be consistent across all FCC-ee energies and is computed as:

\vspace{-0.5em}
\begin{equation*}
\text{K}_{\text{ch}} = \frac{\langle \text{N}_{\text{ch}} \rangle_{\text{pythia}}}{\langle \text{N}_{\text{tot}} \rangle_{\text{pythia}}} \approx 0.472
\end{equation*}

\noindent At parton-level, QCD theory predicts the energy evolution of mean parton multiplicity $\langle \text{n}_{\text{tot}}\rangle$ up to next-to-next-to-next-to-leading order (3NLO) as follows~\cite{Lupia_1998, Dremin_2001}:

\vspace{-0.5em}
\begin{equation}
\langle \text{n}_{\text{tot}} \rangle_{\text{qcd}} \;=\; \frac{y^{-a_1c^2}}{r_0} \ e^{ (2c\sqrt{y} \ +\,A_{\text{NNLO}}(y) + \ A_{\text{3NLO}}(y))}
\end{equation}

\noindent where the NNLO and 3NLO corrections are defined as,
\begin{equation}
A_{\text{NNLO}}(y) = \frac{c}{\sqrt{y}} \Bigg[ r_1 + 2a_2c^2 + \frac{\beta_1}{\beta_0^2}\big(\ln(2y)+2\big) \Bigg]
\end{equation}
\begin{equation}
A_{\text{3NLO}}(y) = \frac{c^2}{y} \Bigg[ a_3c^2 + \frac{r_1^2}{2} + r_2 - a_1\frac{\beta_1}{\beta_0^2}\big(\ln(2y)+1\big) \Bigg]
\end{equation}

\noindent The expansion parameter $y$ in the equations is given as, 
\begin{equation*}
y = \ln \Bigg( \frac{\sqrt{s}}{2Q_0} \Bigg) = \ln \Bigg( \frac{\sqrt{s}}{2\Lambda} \Bigg)
\end{equation*}

\noindent Here, the cut-off scale $Q_0$ is set to the QCD scale $\Lambda$. The remaining coefficients of above equations are defined as,
\begin{equation*}
c=\sqrt{4N_c/\beta_0} \ \ \ \ \ \ \ \beta_0 = 11 - \tfrac{2}{3}N_f \ \ \ \ \ \ \ \beta_1 = 102 - \tfrac{38}{3}N_f
\end{equation*}

\noindent Numerical values of $r_0,r_1,r_2,a_1,a_2,a_3$ are given in~\cite{Capella_2000}. The total parton multiplicity is then translated to the hadron-level variable $\langle \text{N}_{\text{tot}} \rangle_{\text{qcd}}$ through invoking LHPD:

\vspace{-0.5em}
\begin{equation*}
\langle \text{N}_{\text{tot}} \rangle_{\text{qcd}} = \langle \text{n}_{\text{tot}} \rangle_{\text{qcd}} \ \times \ \text{K}_{\text{LHPD}}
\end{equation*}

\noindent The constant $\text{K}_{\text{LHPD}}$ was determined by normalizing the 3NLO theory prediction to the PYTHIA curve at Z-pole:

\vspace{-0.5em}
\begin{equation*}
K_{\text{LHPD}} = 
\frac{\langle N_{\text{ch}} \rangle_{\text{pythia}}(\text{m}_\text{Z})}
     {\langle N_{\text{ch}} \rangle_{\text{qcd}}(\text{m}_\text{Z})} 
     = 0.207
\end{equation*}

\noindent This value closely matches the result from ALEPH~\cite{ALEPH:209gev}. Finally, the neutral correction is applied to the hadron-level theory prediction for comparison with experiment:

\vspace{-0.5em}
\begin{equation*}
\langle \text{N}_{\text{ch}} \rangle_{\text{qcd}} = \langle \text{N}_{\text{tot}} \rangle_{\text{qcd}} \ \times \ \text{K}_{\text{ch}}
\end{equation*}

\begin{figure}[b]
\includegraphics[width=1\linewidth]{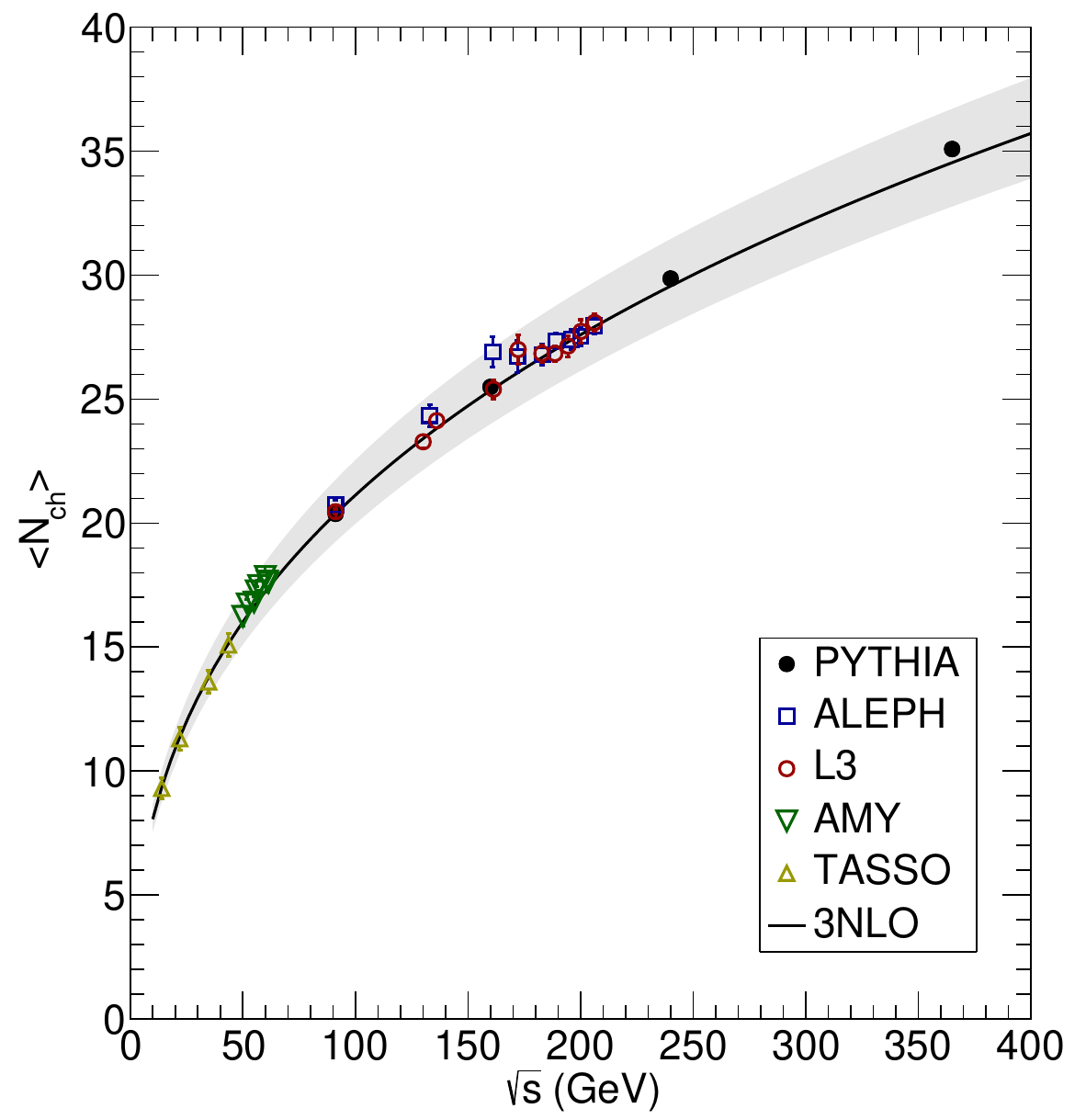}
\caption{\label{fig:13}Energy evolution of $\langle \text{N}_{\text{ch}} \rangle$ obtained in PYTHIA bench-marked with experiment values. The band corresponds to the 3NLO QCD prediction evaluated at $\Lambda=202\pm31$ MeV.}
\end{figure}

\noindent The 3NLO prediction is evaluated at $\Lambda=202\pm31$ MeV (opted from~\cite{ALEPH:209gev}) with $N_f=5,N_c=3$. This curve and uncertainty band is plotted in Fig.~\ref{fig:13} alongside results from PYTHIA and measurements from TASSO~\cite{TASSO:1989}, AMY~\cite{AMY:1990}, L3~\cite{L3:172gev}, and ALEPH~\cite{ALEPH:209gev}. The simulated values accurately follow a logarithmic growth with increasing center-of-mass across the FCC-ee regime, as predicted by QCD. This indicates that perturbative evolution remains reliable well into the multi-hundred-GeV region. The slight deviations emerging at higher energies could possibly hint at the limits of the fixed-order expansion.

\begin{table}[t]
\caption{\label{tab:5} Obtained values of mean charged hadron multiplicities $\langle \text{N}_{\text{ch}}\rangle$ from PYTHIA simulations at FCC-ee energies. }.
\begin{ruledtabular}
\begin{tabular}{ccccc}
Energy & 91.2 GeV & 160 GeV & 240 GeV & 365 GeV \\
\hline
$\langle \text{N}_{\text{ch}} \rangle$ & 20.37 & 25.55 & 26.89 & 27.80 \\
\end{tabular}
\end{ruledtabular}
\end{table}

\clearpage
\subsection{Momentum distribution}
While $\langle \text{N}_{\text{ch}}\rangle$ encapsulates the integrated yield, the inclusive variable $\xi$ offers a differential view of the spread of scaled momentum $x_p$ among particles. It is defined as:

\vspace{-0.5em}
\begin{equation}
\xi \;=\; \ln\!\left(\frac{1}{x_p}\right) 
\qquad 
\text{where,} \ \ \ \ x_p \;=\; \frac{2p}{\sqrt{s}}
\label{eq:xi_def}
\end{equation}

\begin{figure}
\includegraphics[width=1\linewidth]{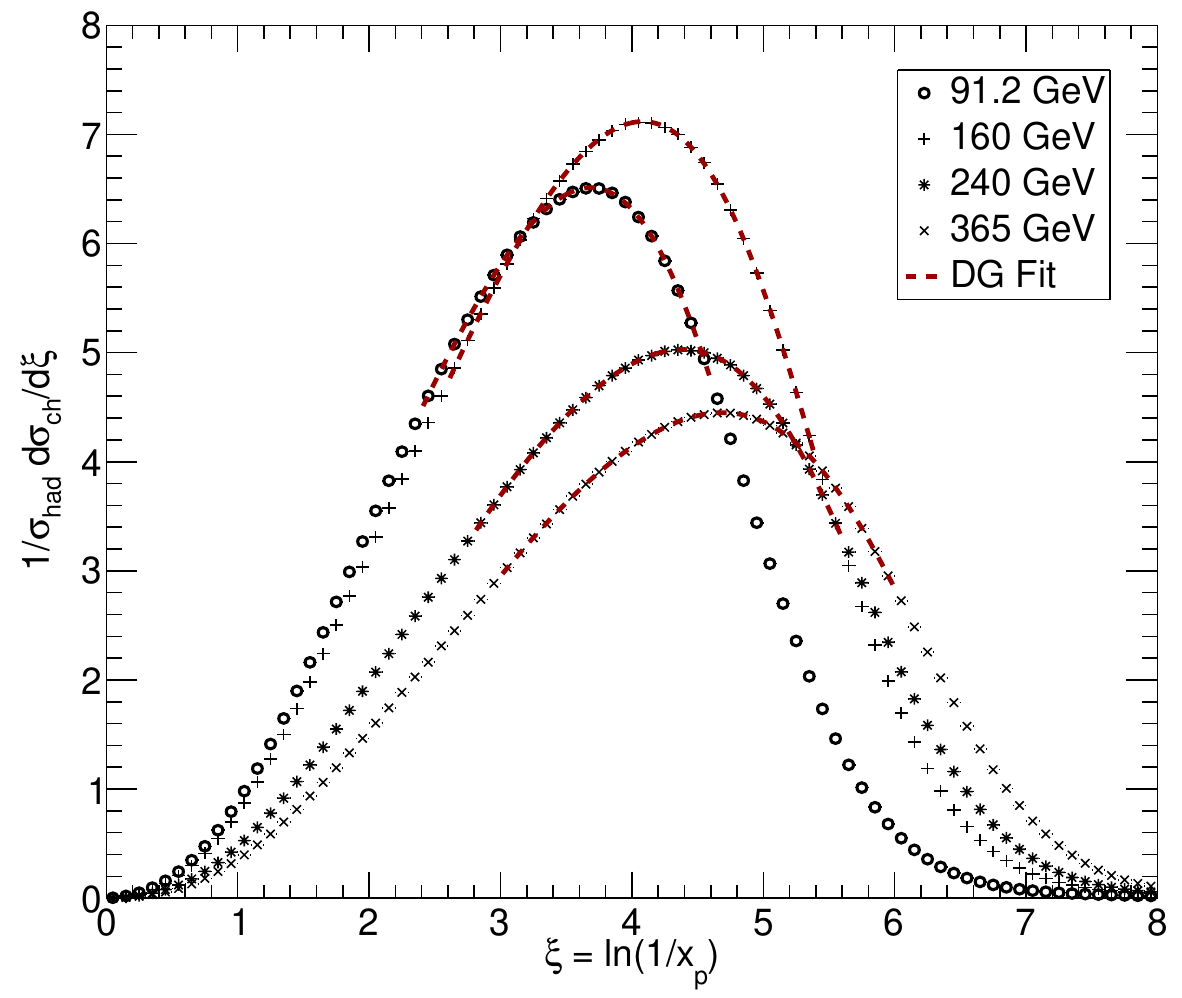}
\caption{\label{fig:14}Spectra of $\xi=ln(1/x_p)$ from PYTHIA at FCC-ee energies and fit results to the distorted Gaussian function.}
\end{figure}

\noindent As partons continue to radiate gluons, many are soft and carry low energy. Interference between successive soft emissions leads to coherent radiation making certain directions become more probable while suppressing others. This coherence results in a gradual narrowing of the parton shower cone, a phenomenon known as angular ordering~\cite{Ermolaev:1981,Dokshitzer:1988}. As a consequence, the $\xi$ spectrum rises in the low-$p$ region, reaches a maximum at intermediate momenta, and then falls off, producing the characteristic hump-backed shape expected by QCD. The central region of this curve can be described either by a Gaussian, a distorted Gaussian (selected for this study), or the limiting spectrum prediction. The simulated $\xi$ distributions and their fits are shown in Fig.~\ref{fig:14}. The Fong-Webber expression~\cite{Fong_1991} is utilized with all five parameters as free variables: normalization, mean, width, skewness, and kurtosis. Fit ranges similar to~\cite{ALEPH:209gev} yield acceptable results and the resulting values are summarized in Table~\ref{tab:6}. The position of the peak of the fitted $\xi$ curve, denoted by $\xi^*$, evolves with energy according to~\cite{Khoze_1997}:

\begin{equation}
\xi^{*} \;=\; y\!\left( \frac{1}{2} + \sqrt{\frac{c_{\xi}}{y}} - \frac{c_{\xi}}{y} \right)
\end{equation}

\noindent where the coefficients $c_{\xi}$ is defined as, 

\begin{equation*}
c_{\xi} = \frac{a^{2}}{16\,N_{c}\,b} \ \ \  \text{where,} \ \ \ a = \frac{11N_{c}}{3} + \frac{2N_{f}}{3N_{c}^{2}} \ \ \ b = \frac{11N_{c}}{3} - \frac{2N_{f}}{3}
\end{equation*}

\begin{figure}[b]
\includegraphics[width=1\linewidth]{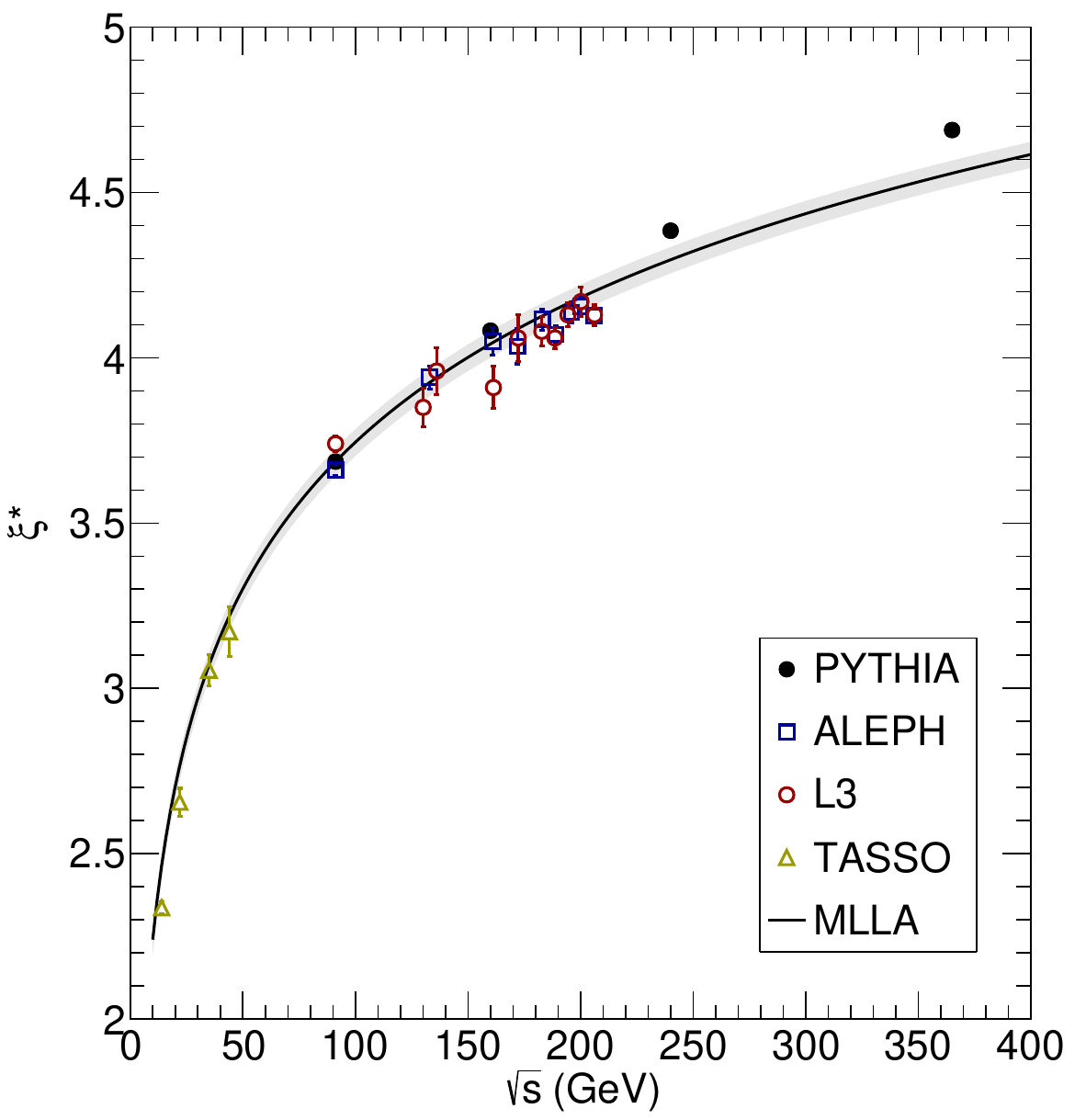}
\caption{\label{fig:15}Energy evolution of $\xi^*$ values determined from the mean of the Gaussian fits to $\xi$ distributions obtained in simulation, bench-marked with results from previous experiments.}
\end{figure}

\noindent Corrections to $\xi^*$ from neutrals is minimal since it does not bring about change to the peak position. The MLLA expression evaluated with $N_f=5,N_c=3$ is normalized to the PYTHIA value at 91.2 GeV and plotted in Fig.~\ref{fig:15}. The uncertainty band corresponds to a variation in $\Lambda=235\pm15$ MeV~\cite{ALEPH:209gev}. The experimental results from TASSO~\cite{TASSO_1990}, L3~\cite{L3:172gev}, and ALEPH~\cite{ALEPH:209gev} mostly follow the theory prediction while simulated values begin to deviate significantly above 160 GeV. The higher $\xi^*$ could be due to parameters in PYTHIA that were tuned at 91.2 GeV not perfectly extrapolating to higher energies. This behavior could also indicate the importance of higher-order logarithmic terms and mass effects not captured in the fixed-order MLLA formulation. Future analyses incorporating next-to-MLLA corrections and dedicated retuning of fragmentation parameters at FCC-ee energies will be essential to refine these predictions and thereby probe the transition between perturbative and non-perturbative QCD domains with greater precision.

\begin{table}[t]
\caption{\label{tab:6} Peak positions $\xi^*$ from fitting the distorted Gaussian function to PYTHIA $\xi$ distributions at FCC-ee energies. }.
\begin{ruledtabular}
\begin{tabular}{cccc}
Energy & Fit range & $\xi^* \ \pm \ \Delta\xi^*_{stat}$ &  $\chi^2/N_{dof}$ \\
\hline
91.2 GeV & 2.4 -- 4.6 & 3.686 $\pm$ 0.087 & 52.5/17
\\
160 GeV & 2.6 -- 5.4 & 4.082 $\pm$ 0.044 & 66.9/23
\\
240 GeV & 2.8 -- 5.6 & 4.384 $\pm$ 0.041 & 33.1/23
\\
365 GeV & 3.0 -- 6.0 & 4.689 $\pm$ 0.037 & 29.0/25
\end{tabular}
\end{ruledtabular}
\end{table}

\clearpage
\section{Conclusions}
An extensive analysis of event shape observables and inclusive hadron spectra at the planned c.m. energies of the FCC-ee has been presented. Samples of $5 \times 10^6$ have been generated in PYTHIA 8.313 at 91.2, 160, 240 and 365 GeV corresponding to the planned FCC-ee program. 

An investigation of event shapes at high center-of-mass energies has been presented. Significant distortions have been observed due to initial state photon radiation (Fig.~\ref{fig:05}) and additional electroweak backgrounds from hadronic decays of Z pairs, W pairs, top-quark pairs, and Higgs (Fig.~\ref{fig:06}-\ref{fig:09}).~Mitigation through radiative cuts (Table~\ref{tab:2}) and subtractive corrections (Table~\ref{tab:3}) will come at a consequential cost of event statistics.~Such strategies must hence be studied further to find an optimal balance.

An extraction of the strong coupling up to NNLO at the four FCC-ee energies has been presented.~The employed fit procedure produces agreeable results (Table~\ref{tab:4}) and the fit ranges used at LEP hold reasonably well (Fig.~\ref{fig:12}).~Since statistical errors will be negligible at the FCC-ee, tight control on systematic uncertainties will be necessary to leverage the improved accuracy of theoretical models.~Hadronisation correction factors computed in PYTHIA showcase a decreasing impact of hadronisation errors with increasing energy (Fig.~\ref{fig:11}).

An examination of inclusive hadron spectra has been presented. The energy evolution of charged hadron multiplicities follows the 3NLO prediction with slight deviations trend emerging above 240 GeV (Fig.~\ref{fig:13}).~The contribution of neutral hadrons to multiplicity is found to be consistent with c.m. energy. Translating parton distributions to the hadron-level by invoking Local Parton-Hadron Duality yields a normalization constant that matches the previous value from ALEPH~\cite{ALEPH:209gev}. The energy evolution of momentum distribution peaks also displays an offset from the MLLA prediction (Fig.~\ref{fig:15}).

The broad scope of the presented study provides several starting points for future extensions. Immediate additions could incorporate additional event shape observables~\cite{Kardos_2021}, other Monte Carlo event generators, resummed calculations~\cite{Dissertori_2009}, higher fixed-order calculations~\cite{Kardos_2021}, and corrections for mass effects of heavy quarks. Combining simulated results with the DELPHES framework~\cite{DELPHES} can delve deeper into understanding the influences of detector effects on systematic uncertainties. 

\begin{acknowledgments}
\noindent The authors thank Andrii Verbytskyi, MPI Physics, for his inputs regarding the scope of this work, and Otmar Biebel, LMU Physics, for discussions regarding his TESLA collider studies. The author PM thanks Davide Aguglia, CERN SY-EPC, for encouraging the project.
\end{acknowledgments}
\bibliography{FCC.bib}
\end{document}